\documentstyle[12pt]{article}

\begin{document}

\thispagestyle{empty}

\hfill \parbox{45mm}{{ECT*-99-8} \par June 1999}

\vspace*{5mm}

\begin{center}
{\LARGE Virtual dipoles and large fluctuations}

\smallskip

{\LARGE in quantum gravity}

\vspace{12mm}

{\large Giovanni Modanese}
\footnote{e-mail: modanese@science.unitn.it}

\medskip

{\em European Centre for Theoretical Studies in Nuclear Physics
and Related Areas \par
Villa Tambosi, Strada della Tabarelle 286 \par
I-38050 Villazzano (TN) - Italy}

{and}

\medskip

{\em I.N.F.N.\ -- Gruppo Collegato di Trento \par
Dipartimento di Fisica dell'Universit\`a \par
I-38050 Povo (TN) - Italy}
	
\end{center}

\vspace*{5mm}

\begin{abstract}

The positive energy theorem precludes the possibility of Minkowski flat
space decaying by any mechanism. In certain circumstances, however, large
quantum fluctuations of the gravitational field could arise---not only at
the Planck scale, but also at larger scales. This is because there exists a
set of localised weak field configurations which satisfy the condition
$\int d^4x \sqrt{g}R = 0$ and thus give a null contribution to the Einstein
action. Such configurations can be constructed by solving Einstein field
equations with unphysical dipolar sources. We discuss this mechanism and
its modification in the presence of a cosmological term and/or an external
field.

\medskip
\noindent 04.20.-q Classical general relativity.

\noindent 04.60.-m Quantum gravity.

\medskip
\noindent Key words: General Relativity, Quantum Gravity

\bigskip

\end{abstract}

\subsection*{1. Is Minkowski space stable?}
 
Although flat spacetime is a trivial solution of the vacuum Einstein
equations, the {\it stability} of this solution was questioned for a long
time, mainly because in General Relativity (unlike in electrodynamics and
other field theories) it is impossible to give a positive semidefinite
expression for the field energy density. The numerous approaches to this
problem reported in the literature can be essentially classified in the
context of classical field theory, thermal field theory and quantum field
theory. Let us quote very shortly some of the known facts. 
 
(1) {\it Classical field theory} - The long-outstanding conjecture that the
total energy (A.D.M.\ energy) of asymptotically flat manifolds is positive
semidefinite and that only Minkowski space has zero energy was finally
proven by Schoen and Yau in 1979 \cite{r1} and is now known as ``the
positive energy theorem". Since energy is conserved, this theorem seems to
preclude the possibility of flat space decaying by any mechanism. 
 
(2) {\it Thermal field theory} - The instabilities of Minkowski space at a
finite temperature were studied by Gross, Perry and Yaffe in 1982
\cite{r2}. They concluded that in hot flat space there are two distinct
sources of instability: the large-wavelength density fluctuations of the
thermal gravitons and the nucleation of black holes. 
 
(3) {\it Quantum field theory} - One can consider, at least formally, the
functional integral of the theory, which represents a sum over all
possible field configurations weighed with the factor $\exp[i\hbar
S] =\exp \left[ \frac{-i\hbar}{16\pi G} \int d^4x
\sqrt{g(x)}R(x) \right]$ and possibly
with a factor due to the integration measure. The Minkowski space is a
stationary point of the vacuum action and has maximum probability. 
``Off-shell" configurations, which are not solutions of the vacuum Einstein
equations, are admitted in the functional integration but are strongly
suppressed by the oscillations of $e^{i\hbar S}$.
 
In this letter (based in part upon our work \cite{r3})  we shall be
concerned with the quantum case. Even though any tunnelling to a ground
state different from Minkowski space appears to be impossible due to the
positive energy theorem, we shall see that the quantum fluctuations about
flat space can be enhanced in certain conditions.

\subsection*{2. The quantum of curvature fluctuation}

Due to the appearance of the dimensional constant $G$ in the Einstein
action, the most probable quantum fluctuations of the gravitational field
grow at very short distances, of the order of $L_{Planck}= \sqrt{G
\hbar/c^3} \sim 10^{-33} \ cm$. This led Hawking, Coleman and
others to depict spacetime at the Planck scale as a ``quantum foam"
\cite{r4}, with high curvature and variable topology. 
 
Let us reformulate this argument in short. Suppose we start with a flat
configuration, then a curvature fluctuation appears in a region of size
$d$.  How much can the fluctuation grow before it is suppressed by the
oscillating factor $e^{iS}$? (We set $\hbar=1$ and
$c=1$ in the following.) The contribution of the fluctuation to the action
is of order $Rd^4$; both for positive and for negative $R$, the
fluctuation is suppressed when this contribution exceeds $\sim 1$ in
absolute value, therefore $|R|$ cannot exceed $\sim G/d^4$. This means
that the fluctuations of $R$ are stronger at short distances---down to
$L_{Planck}$, the minimum physical distance. 
 
Clearly if the curvature is large, then the metric is locally far from
being flat and the factor $\sqrt{g}$ in the action is not trivially
$\sim 1$, thus the estimate above is only approximate. In the following,
however, we shall focus on the case of a weak field with small curvature,
at distances much larger than $L_{Planck}$ (without any topology change).
In this case we could say that $R_0 \equiv G/d^4$ represents the
``quantum of curvature fluctuation", very small at macroscopic scale. As
shown above, the number of such quanta in a certain region does not exceed
$N \sim 1$; therefore fluctuations are practically irrelevant and
spacetime looks almost perfectly flat at distances much larger than
$L_{Planck}$.

\subsection*{3. Virtual dipoles containing $N$ +/quanta and $N$ --/quanta}

At this point one might rise the following objection. Suppose two quanta
of curvature fluctuation with opposite signs pop up in flat space, in two
adjacent regions having the same size $d$. This does not change the total
action. Then the negative fluctuation can grow up to comprise 2, 3, ...
$N$ quanta, provided the same happens with the positive fluctuation,
because the total action of a configuration containing $N$ +/quanta and
$N$ --/quanta is the same as the flat space action. Can this represent a
possible instability, a way for the gravitational field to ``run away" from
the flat configuration, causing strong fluctuations of the metric also at
scales larger than Planck scale? 
 
This idea might seem naive and qualitative, and perhaps every beginner in
General Relativity had it for a minute. However, it can be precised and
made more rigorous. It is possible to construct explicitly field
configurations with the property above, namely having scalar curvature
which vanishes identically almost everywhere, except in two adjacent
regions -- one with positive $R$ and the other with negative $R$ -- in
such a way that the total integral of $\sqrt{g}R$ is zero. 
 
One can regard each of these field configurations as the field generated
by a virtual ``mass dipole". In fact, they can be defined as the solutions
of the Einstein field equations with a dipolar source---a positive and a
negative mass with certain sizes, chosen in such a way that the total
integral of the scalar curvature is zero. Such sources are clearly
unphysical and do not exist in the real world; however, here we are not
interested into a solution of the classical field equations with physical
sources but into any field configuration which can cause strong
fluctuations in the functional integral. 
 
Let us consider a solution $g_{\mu \nu}(x)$ of the Einstein equations
	\begin{equation} R_{\mu \nu}(x) - \frac{1}{2} g_{\mu \nu}(x) R(x)
= -8 \pi G T_{\mu \nu}(x), 
\label{e1} 
\end{equation}
	with a (covariantly conserved) source $T_{\mu \nu}(x)$ obeying the
additional integral condition
	\begin{equation} 
\int d^4x \sqrt{g(x)} {\rm Tr} \, T(x) = 0.  
\label{e2}
\end{equation}
	Taking into account the trace of eq.\ (\ref{e1}), namely $R(x)=8
\pi G {\rm Tr} \, T(x)$, we see that the action $\int d^4x \sqrt{g}R$ 
computed for this solution is zero.
 
As an example of an unphysical source which satisfies (\ref{e2}) one can
consider a static dipole centred at the origin ($m,m'>0$): 
	\begin{equation} 
T_{\mu \nu}({\bf x}) = \delta_{\mu 0} \delta_{\nu 0}
\left[m f({\bf x}+{\bf a}) - m' f({\bf x}-{\bf a}) \right].
\label{e3}
\end{equation} 
	Here $f({\bf x})$ is a smooth test function centred at ${\bf
x}=0$, rapidly decreasing and normalised to 1, which represents the mass
density.  The range of $f$, say $r_0$, is such that $a \gg r_0 \gg
r_{Schw}$, where $r_{Schw}$ is the Schwarzschild radius corresponding to
the mass $m$. The mass $m'$ is in general different from $m$ and chosen in
such a way to compensate the small difference, due to the
$\sqrt{g}$ factor, between the integrals
	\begin{equation}
I^+ = \int d^3x \sqrt{g({\bf x})} f({\bf x}+{\bf a}) \ \ \ {\rm and} \ \ \
I^- = \int d^3x \sqrt{g({\bf x})} f({\bf x}-{\bf a}).
\label{e4}
\end{equation}

The procedure for the construction of the desired field
configuration is the following. One first considers Einstein equations
with the source (\ref{e3}). Then one solves them with a suitable method,
for instance in the weak field approximation. Finally, knowing
$\sqrt{g(x)}$ one computes the two integrals (\ref{e4}) and adjusts the
parameter $m'$ in such a way that
	\begin{equation} 
mI^+ - m'I^- = 0
\label{e5} 
\end{equation} 
	(see \cite{r3}, where these configurations were called ``zero modes
of the Einstein action"). To first order in $G$, the relation between $m$
and $m'$ turns out to be
	\begin{equation} 
m' \simeq m(1+8\pi Gm).
\label{e6}
\end{equation} 
	Note that these field configurations are in no way singular, so we
do not expect them to be suppressed by the functional integration measure.

\subsection*{4. Does the effective gravitational lagrangian contain a
scale-dependent cosmological term?}

What can stabilise the Einstein action with respect to the dipole
fluctuations described above? Possibly an additional term in the
lagrangian. $R^2$ terms are only relevant at very short distances, however,
which is not the case here. Another typical addition is the cosmological
term $(\Lambda/8\pi G) \int d^4x \sqrt{g}$. It is immediate to check, using
eq.s (\ref{e4})--(\ref{e6})  above, that such a term receives
non-vanishing contributions from dipole fluctuations, and will therefore
suppress them.
 
It is known that $\Lambda$, if not zero, is very small in our
universe;  its effective value, however, could depend on the scale, being
very small at cosmological distances but somewhat larger at short
distances. 
 
This concept, namely that the gravitational lagrangian may comprise a
scale-dependent cosmological term, originally emerged in the Euclidean
theory of gravity on the Regge lattice. Recent non-perturbative numerical
simulations \cite{r5} allow to study a ``discretized spacetime" whose
dynamics is governed by an action containing $G$ and $\Lambda$ as bare
parameters; it turns out that as the continuum limit is approached, the
adimensional product $|\Lambda_{eff}|G_{eff}$ behaves like
	\begin{equation} 
|\Lambda_{eff}|G_{eff} \sim (l_0/l)^\gamma
\label{e7}
\end{equation} 
	where $l$ is the scale, $l_0$ is the lattice spacing, $\gamma$ a
critical exponent and the sign of $\Lambda_{eff}$ is negative (for an
earlier discussion see \cite{r6}). Furthermore, one can reasonably assume
that $l_0 \sim L_{Planck}$, and that the
scale dependence of $G_{eff}$ is much weaker than that of $\Lambda_{eff}$.
 
A scale dependence of $\Lambda_{eff}$ like that in eq.\ (\ref{e7}) also
implies that any bare value of $\Lambda$, expressing the energy density
associated to the vacuum fluctuations of the quantum fields including the
gravitational field itself, is ``relaxed to zero" at long distances just by
virtue of the gravitational dynamics, without any need of a fine tuning.
One would have, in other words, a purely gravitational solution of the
cosmological constant paradox.
 
We do not intend to discuss here whether (and for what values of $\gamma$)
an {\it ansatz} like eq.\ (\ref{e7}), with $\Lambda_{eff}<0$, is
compatible with the most recent estimates of the Hubble parameter
\cite{r7}. In fact, admitted that $\Lambda_{eff}$ depends on the scale
$l$, it is clear that the determination of this dependence and a
comparison with the observational constraints is a complex problem, given
the enormous range spanned by $l$. One can consider at least four domains:
a cosmological scale, a scale of the order of the solar system size
(compare \cite{r8}), a laboratory scale (in a wide sense, i.e., down to
the subnuclear and $GeV$ scale \cite{r9}) and the Planck scale.
While in this work we are mostly concerned with the laboratory scale, we
shall keep a general approach and just denote by $\Lambda_{eff}(l)$ the
general unknown function which gives the scale dependence of
$\Lambda_{eff}$.

\subsection*{5. Coupling to an external field}

We have seen that a scale-dependent cosmological term in the effective 
gravitational action is able to suppress the virtual dipole
fluctuations. The actual existence of this term can be regarded as a
consistent hypothesis, suggested by some results of lattice theory. But
how can we check that the whole idea makes sense, i.e.\ that on one hand
``dangerous" dipole fluctuations are admitted by the Einstein action and on
the other hand an intrinsic cosmological term is there to suppress
them? 
 
We can imagine a situation in which this latter term is canceled, in some
region of spacetime, by coupling gravity to a suitable {\it external}
field $F(x,t)$. Then, if virtual dipoles really exist, they will be free
to grow in this region and cause abnormally large fluctuations. The
amplitude of these fluctuations cannot be predicted at this stage, but
eventually they could lead to a sort of partial ``thermalization" of the
gravitational field in this region. 
 
The function $F(x,t)$ represents an assigned classical field, not a
variable of the functional integral. It couples to gravity
through its energy-momentum tensor
	\begin{equation} 
T_{\mu \nu} = \Pi_{\mu} \partial_{\nu} F - g_{\mu \nu} {\cal L}
\label{e8}
\end{equation} 
	where $\Pi_{\mu}$ is the canonically conjugated momentum $\delta
{\cal L}/\delta(\partial^{\mu}F)$. For instance, for a classical scalar
field $\phi$ (typically describing, in the context of quantum field
theory, coherent matter of some kind) we have 
	\begin{eqnarray}
& & {\cal L} =\frac{1}{2}
\left( \partial_\alpha \phi \partial^\alpha \phi - m^2 \phi^2 \right), \\
& & \Pi_{\mu}=\partial_{\mu} \phi, \\
& & T_{\mu \nu}=\partial_{\mu} \phi
\partial_{\nu} \phi - g_{\mu \nu} {\cal L}
\end{eqnarray}
	and the interaction term in the action takes the form
	\begin{equation} 
S' = 8\pi G \int d^4x \, h_{\mu \nu} T^{\mu \nu} = 8\pi G \int d^4x
(h_{\mu \nu} \partial^{\mu} \phi \partial^{\nu} \phi - {\rm Tr} 
h \, {\cal L}).
\label{e9}
\end{equation} 
	The term $h_{\mu \nu} \partial^{\mu} \phi \partial^{\nu} \phi$
can be regarded as a source term for $h_{\mu \nu}$, while the term
$\int d^4x {\rm Tr} h \, {\cal L}$ interferes with the cosmological term: 
we recall that, to first order in $h$,
$\sqrt{g} \simeq 1+\frac{1}{2} {\rm Tr} h$;  therefore, if at some
point $x$ the condition
	\begin{equation} 
\frac{\Lambda_{eff}(a)}{8\pi G} + {\cal L}(\phi(x)) = 0
\label{e10}
\end{equation} 
	is satisfied, at that point large virtual dipoles fluctuations of
scale $a$ can appear. 

Eq.\ (\ref{e10}) can be regarded as a parametric
equation for $a(x)$:  given the value of the coherent lagrangian density
at $x$, one finds a corresponding value for $\Lambda_{eff}$ and thus for
the scale $a$ of the dipole fluctuations allowed at $x$. (This makes 
sense if $\phi$ varies on a scale much larger than $a$.
Also note that being
$\phi$ an assigned ``off shell" field, ${\cal L}(\phi(x))$ is not
necessarily an extremal value.) The scale obtained in this way can turn
out to be physically relevant, or not---particularly if very short.

\subsection*{6. Conclusions}

This work aims at pointing out a peculiar quantum behaviour of the
gravitational field which is usually disregarded, but cannot a priori be
excluded. The possibility of anomalous growth of the ``dipole fluctuations" 
described above appears to be an intrinsic property of the Einstein
action.
 
It is therefore important to understand to what extent these fluctuations
are affected by a cosmological term in the action or by the coupling to an
external field---which breaks the translation symmetry of Minkowski
spacetime and could thus trigger the fluctuations in certain regions.
 
Our analysis is limited by the poor present knowledge of the quantum
dynamics of the gravitational field; in particular, it seems impossible to
predict at this stage the exact amplitude of the dipole fluctuations and 
the scale dependence $\Lambda_{eff}(l)$ (if any).

\bigskip \bigskip

This work has been partially supported by the A.S.P.\ -- Associazione per lo 
Sviluppo Scientifico e Tecnologico del Piemonte -- Turin -- Italy.

\end{document}